\documentclass[prl,twocolumn,showpacs,nofootinbib]{revtex4}
\usepackage{amssymb}
\usepackage{amsmath}
\usepackage{epsf}
\usepackage{mathrsfs}
\usepackage[dvips]{graphics}
\usepackage[dvips,usenames]{color}
\begin{document}
\def \beq{\begin{equation}}
\def \eeq{\end{equation}}
\def \bea{\begin{eqnarray}}
\def \eea{\end{eqnarray}}
\def \bem{\begin{displaymath}}
\def \eem{\end{displaymath}}
\def \P{\Psi}
\def \Pd{|\Psi(\boldsymbol{r})|}
\def \Pds{|\Psi^{\ast}(\boldsymbol{r})|}
\def \Po{\overline{\Psi}}
\def \bs{\boldsymbol}
\def \bl{\overline{\boldsymbol{l}}}
\def \bR{\overline{R}}
\title{Bogomol'nyi bound and 
screw dislocations in a mesoscopic smectic-A}
\author{Eric Akkermans, Sankalpa Ghosh and Amos Schtalheim}
\affiliation{Department of Physics, Technion Israel Institute of 
Technology,
32000 Haifa, Israel}
\date{\today}
\begin{abstract}
The de Gennes free energy functional of an infinite smectic-A liquid 
crystal at the dual point is shown to be topological and to depend only on the number of screw dislocations and the anisotropy. 
This result generalizes the existence of a Bogomol'nyi bound to an anisotropic system. The role of the boundary of a finite mesoscopic smectic is to provide a mechanism for the existence of thermodynamically stable screw dislocations. We obtain a closed expression for the corresponding free energy and a relation between the applied twist and the number of screw dislocations. 
\end{abstract}

\maketitle
A successful Êand striking analogy between different phases of a liquid crystal and those of 
a superconductor has been proposed by de Gennes back in 1972 
\cite{DG72}. It provides an example of how geometric and topological
considerations lead to similar behaviours in two different physical
systems \cite{Kamien02}. 
This analogy has been further used by Renn and Lubensky to predict a new phase 
called the Twist Grain Boundary (TGB) phase \cite{RL88}, whose existence has been experimentally confirmed 
\cite{Good88, GS90, KIhn92, NPGN93, NPGN98, RPRYN01, ASSTS03}. This phase is the liquid crystal analogue of 
the Abrikosov vortex lattice. It consists of screw dislocations
that are topological defects. The TGB phase lies 
in between the smectic-A (SmA) and the cholesteric phase. In this letter, we show that 
the de Gennes free energy of a SmA has an integrability point at which  it is topological. This leads to the existence of screw dislocation solutions, 
whose thermodynamic stability is obtained in the mesoscopic limit yet to be specified and for the geometry of an infinitely long cylinder of 
radius $R$. We characterize the smectic-A phase by the following 
set of length scales: the transverse and the longitudinal coherence lengths 
$\xi_{\perp}$ and $\xi_{\parallel}$, the
twist penetration depth $\lambda_{T}$,  and the wavenumber $q_s=\frac{2\pi}{d}$ of the 
density modulation of the nematogen 
molecules along the $z$-axis of the cylinder. The cholesteric pitch length defined by $L_p=
\frac{2\pi}{k_0}$ \cite{RL88} is generally larger than $\xi_{\perp}$. Here $d$ is the separation between smectic layers and
$k_0$ is the wavenumber of the Frank director in the cholesteric phase.

We first show that when written in a  dimensionless Êform, the de Gennes free energy density ${\cal F}$ of an effective two-dimensional system is solely characterized by two
dimensionless parameters, namely the twist Ginzburg parameter $\kappa=
\frac{\lambda_{T}}{\xi_{\perp}}$ and the anisotropy parameter $\alpha=
\sqrt{2} \kappa \xi_{\parallel} q_s \propto \xi_{\parallel}/\xi_{\perp}$. The dual point \cite{SJ69} is defined by $\kappa = 1/\sqrt{2}$. At this point and for a large  system, the de Gennes free energy in the presence of a finite twist is a topological quantity that can be written
\beq \int {\cal F} = 2\pi l(1-\alpha^2), 
\label{inenergy}\eeq
where the integer $l$ accounts for the number of screw dislocations. This expression corresponds to the Bogomol'nyi bound \cite{Bogo77} for the liquid crystal. However, from (\ref{inenergy}) it appears that there is no mechanism to 
select the number $l$ of defects, 
and the free energy is minimum for $l=0$. 
A way to thermodynamically stabilize solutions with $l \ne 0$ is to consider 
a mesoscopic finite size smectic, 
just like the unconventional vortex patterns obtained in mesoscopic 
superconductors \cite{Geim1}. 
A mesoscopic smectic-A for the geometry of a cylinder of radius $R$ is characterized by $R \ll  
L_p$. Moreover, for $R > \xi_{\perp}$, we show that for a wide range of parameters $k_0$ and $\alpha$, a mesoscopic SmA Êcan accommodate screw dislocations and the selection mechanism for the equilibrium value of $l \ne 0$ is provided by the boundary conditions imposed on the total twist through the sample, in a way quite similar to the Little-Parks setup in superconducting rings. 

We first Êconsider some of the features of the SmA phase and the conditions of Êits description through the 
de Gennes free energy \cite{DG72}. The order parameter that characterizes the SmA phase 
is the first harmonic of the density of the nematogen molecules. It can be written as $\P(\bs{r})=|\P| e^{i \Phi}$ where 
$\Phi=  q_s \bs{n} \cdot \bs{r}$. The unit vector $\bs{n}$ 
is the Frank director that specifies the molecular 
alignment. For a perfect SmA phase we have $\bs{n}=\{0,0,1\}$.
The smectic planes are defined by $z= ld, (l \in \mathbb{Z})$, so that the 
phase $\Phi=2\pi l$ \cite{DGP93,CL,luben1}.  A screw dislocation is a spiral staircase structure of the smectic planes \cite{luben1,CL} 
characterized by a finite winding number. The free energy of a SmA is given by the sum of the Landau ($F_L$) and the Frank 
($F_F$) contributions \cite{DGP93} which are, respectively: 
\beq F_{L}=\int r|\P|^2 + \frac{g}{2}|\P|^4 + C_{\parallel}
|\frac{\partial \P}{\partial z}|^2 + C_{\perp}
\left(|\frac{\partial \P}{\partial x}|^2 + |\frac{\partial \P}{\partial y}|^2\right) 
\label{BGLE}\eeq
\beq F_{F}=\frac{1}{2} \int 
K_1 (\bs{\nabla}\cdot \bs{n})^2 + K_2 (\bs{n}.\bs{\nabla} \times \bs{n})^2
+ K_3(\bs{n} \times \bs{\nabla} \times \bs{n})^2. \nonumber \eeq
The Frank term $F_F$ is the sum of energies associated to bend, twist and 
splay of $\bs{n}$. For a perfect SmA, the Frank energy vanishes.
Hereafter we consider only the twist part in $F_F$. 
The de Gennes free energy \cite{DG72,DGP93,LBG98} 
is obtained as the limit of the total free energy $F = F_L + F_F$, 
when the Frank director $\bs{n}$ slightly deviates from the $z$-axis, namely $n_z \approx 1$
and both $n_{x}$ and $n_{y}$ are much 
smaller than $n_{z}$. This approximation together with the condition that $R \ll  L_p$ defines the mesoscopic limit. The de Gennes free energy is thus given by \cite{DGP93}
\bea F_{dG} &=& \int r|\P|^2 + \frac{g}{2}|\P|^4 + 
C_{\parallel}
|\frac{\partial \P}{\partial z}|^2 Ê Ê\\
&+ & C_{\perp}
|(\bs{\nabla}_{\perp} - iq_s \delta \bs{n}_{\perp})\P|^2 + 
\frac{K_2}{2}(\bs{\nabla}_{\perp} \times \delta\bs{n}_{\perp})^2 Ê\nonumber  \label{CVE1}, \eea 
where $\bs{\nabla}_{\perp}=\{\partial_x,\partial_y\}$. 
We have introduced the two-dimensional
vector $\delta \bs{n}_{\perp}=\bs{n}-\hat{\bs{z}}$. 
Since ${\bs n}^2=1$, at the lowest order $\delta \bs{n}_{\perp}$ 
lies in the $x$-$y$ plane.  The lengths $\xi_{\perp}$ and $\xi_{\parallel}$ over which the SmA 
order parameter changes Êrespectively 
in the $x$-$y$ plane and along the $z$-axis are obtained by comparing the corresponding gradient 
terms with the first term in the energy functional (3). This 
sets $\xi_{\perp}=\sqrt{C_{\perp}/ |r|}$ and 
$\xi_{\parallel}=\sqrt{C_{\parallel} / |r|}$. The 
twist penetration depth Êis 
$\lambda_{T} = \sqrt{\frac{g K_2}{C_{\perp} q_s^2 |r|}}$ \cite{DG72, RL88}.
In the mean field approximation, 
only the coefficient $r$ depends on temperature \cite{DG72,RL88, LBG98}, 
so that the Ginzburg parameter 
$\kappa=\frac{\lambda_T}{\xi_{\perp}}$ is temperature independent. 
By rescaling all the lengths in units of 
$\sqrt{2} \lambda_T$, namely defining the scaling relations
$\frac{r^2}{2g}{\cal F} = F_{dG}$, 
$\frac{-r}{g}|f|^2 = |\P|^2$, $\sqrt{2}\lambda_T\overline{x}= x$, 
$\sqrt{2}\lambda_T q_s=\overline{q}_s$ and $\overline{\bs{\nabla}}_{\perp}=
\sqrt{2}\lambda_T {\bs{\nabla}}_{\perp}$, the de Gennes free energy (3) can be rewritten, up to a constant, in the dimensionless form 
\bea {\cal F}&=& \kappa^2(|f|^2 -1)^2 + \alpha^2|f|^2 + \nonumber \\
&+& |(\overline{\bs{\nabla}}_{\perp} - i\overline{q}_s \delta\bs{n}_{\perp})f|^2 + 
\frac{1}{2}(\overline{\bs{\nabla}}_{\perp} \times \overline{q}_s\delta \bs{n}_{\perp})^2, 
\label{dlfreeen1}\eea 
where both $f$ and $\delta \bs{n}_{\perp}$ do not depend on $z$. 
The anisotropy parameter $\alpha^2 = 2\kappa^2 q_s^2 
\xi_{\parallel}^2 = \overline{q}^2_s C_{\parallel}/C_{\perp}$ contains $|r|$ and it Êis therefore  temperature dependent. The expression (\ref{dlfreeen1}) is analogous to the dimensionless Ginzburg-Landau free energy density of a two-dimensional superconductor in a magnetic field \cite{dgsuper,eric99}, 
except for the additional term $ \alpha^2|f|^2 $. This term is proportional to $1/ d^2$, and it accounts for the interaction between the nematogen density in SmA layers. 
The equilibrium equations are obtained by minimizing the functional (\ref{dlfreeen1})
with respect to $f^*$ and $\delta\bs{n}_{\perp}$. It gives  
\bea |\overline{\bs{\nabla}}_{\perp} - iq_{s}\delta\bs{n}_{\perp}|^2 f &=& 
2\kappa^2 f(1-|f|^2) - \alpha^2 f \nonumber \\
\overline{\bs{\nabla}}_{\perp} \times (\overline{\bs{\nabla}}_{\perp} \times 
\overline{q}_s \delta \bs{n}_{\perp})&=&2\bs{j}. \label{MA}\eea 
The second relation is the liquid crystal 
analogue of the Maxwell-Amp\`ere equation. It is written in terms of the two-dimensional liquid crystal current density, defined by 
$\bs{j} = Im(f^{\ast} \overline{\bs{\nabla}}_{\perp} f) - |f|^2 
\overline{q}_s\delta\bs{n}_{\perp}$. 
This current density vanishes for a perfect smectic. 
Using the phase of the SmA order parameter, it can also be written as 
$\bs{j}=|f|^2 (\overline{\bs{\nabla}}_{\perp} 
\Phi - \overline{q}_s \delta\bs{n}_{\perp})$. In a superconductor, 
the vanishing of the in-plane circulation of the current density 
around a closed contour $\Gamma$ implies the quantization of the magnetic flux. 
Similarly, in a SmA the vanishing of the circulation of the 
two-dimensional current density $\bs{j}$ around a closed contour implies that the smectic 
planes are lifted by an integer multiple of the layer separation $d$. We define Êthe liquid crystal analogue of the London fluxoid Êby
\beq \oint_{\Gamma} 
\overline{\bs{\nabla}}_{\perp}\Phi.d\overline{\bs{l}} = Ê
\oint_{\Gamma} (\frac{\bs{j}}{|f|^2} + \overline{q}_s \delta \bs{n}_{\perp}).d\overline{\bs{l}} = 2 \pi l, Ê\label{fluxoid}\eeq
where $l$ is an integer. We now make use of the following identity which holds 
in two dimensions only
\beq |(\overline{\bs{\nabla}}_{\perp} - i\overline{q}_{s}\delta \bs{n}_{\perp})f|^2
=|\overline{{\cal D}}f|^2 + (\overline{\bs{\nabla}}_{\perp} \times \bs{j})+ 
|\overline{\bs{\nabla}}_{\perp} \times 
\overline{q}_s \delta \bs{n}_{\perp}||f|^2.\label{BGI}\eeq 
The scalar operator $\overline{{\cal D}}$ is defined by 
$\overline{{\cal D}}=\partial_{\overline{x}} + i\partial
_{\overline{y}} -i\overline{q}_s(\delta n_x 
+ i \delta n_y)$. We denote by $\Omega$ the cross-sectional area of the cylinder and by $\partial \Omega$ its boundary. We set boundary conditions by imposing 
that the system behaves as a perfect smectic 
at large distances, namely $|f| \rightarrow 1$ and $\bs{j} 
\rightarrow 0$ on $\partial \Omega$. Making use of those boundary conditions and of the identity (\ref{BGI}), the free energy (\ref{dlfreeen1}) at the dual point $\kappa = 1/ \sqrt{2}$ can be written, up to a constant, as
\begin{eqnarray}
\int_{\Omega} {\cal F} &=& \int_{\Omega} {1 \over 2} \left( 1 - |f|^2 - \alpha^2 -  |\overline{\bs{\nabla}}_{\perp} \times \overline{q}_s \delta \bs{n}_{\perp}| \right)^2 + |\overline{{\cal D}}f|^2 \nonumber \\ 
&+& (1 - \alpha^2) \oint_{{\partial \Omega}} (\frac{\bs{j}}{|f|^2} + \overline{q}_s \delta \bs{n}_{\perp}).d\overline{\bs{l}}. \label{dggbogo}
\end{eqnarray}
The boundary integral results from both the Stokes theorem and the boundary conditions. 
Making use of the relation (\ref{fluxoid}), it is reduced to the constant term 
$2 \pi l (1 - \alpha^2)$. Then the free energy (\ref{dggbogo}) is reduced to the sum of two positive terms. It is minimum when each of them vanishes, so that the minimum free energy is given by the so-called Bogomol'nyi bound (\ref{inenergy}) and the corresponding field configurations (\ref{MA}) are now solutions of the Bogomol'nyi like equations \cite{Bogo77}
\beq
\overline{{\cal D}}f = 0,~ \mbox{and} ~
|\overline{\bs{\nabla}}_{\perp} \times \overline{q}_s \delta \bs{n}_{\perp}| + \alpha^2 = 1 - |f|^2. \label{bogo}\eeq 
It is important at this point to notice the role played by the nonlinear term in the free energy (\ref{dlfreeen1}) in order to reach the Bogomol'nyi bound. The second equation in (\ref{bogo}) expresses the dual relation between the twist and 
$|f|$. By eliminating the twist between these two equations, we obtain for $|f|$ the nonlinear equation 
$\overline{\nabla}_{\perp}^2 ln|f|^2 = 2(\alpha^2+ |f|^2 -1)$. 
For $\alpha=0$, it is a Liouville 
equation which is known \cite{Liouville} to admit families of vortex solutions characterized by the winding number $l$ that appears in the free energy (\ref{inenergy}). The existence of a Bogomol'nyi bound for the de Gennes free energy of a SmA and the equations (\ref{bogo}) are one of the main results of this letter. 

\begin{figure}[h]
\begin{center}
\vspace*{11pt}
\leavevmode \epsfxsize0.75\columnwidth
\epsfbox{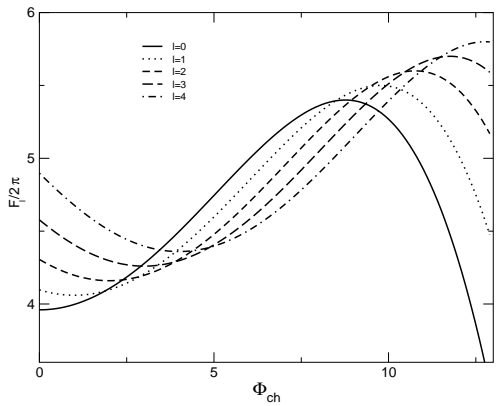}\vskip1.0pc
\caption{\it Free energy $F_l$ of a mesoscopic SmA as a function of the twist $\Phi_{ch}$ as given by (\ref{r14}) with $\overline{R} = 8$ and $\alpha^2 = 0.9$. 
}
\label{fig3}
\end{center}
\end{figure}

\begin{figure}[h]
\begin{center}
\vspace*{11pt}
\leavevmode \epsfxsize0.75\columnwidth
\epsfbox{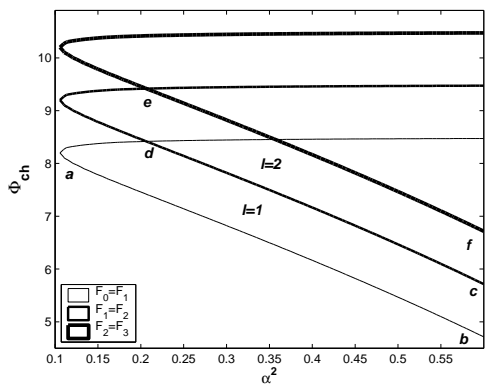}\vskip1.0pc
\caption
{\it Phase diagram in the $\Phi_{ch}$-$\alpha^2$ plane, of a mesoscopic smectic-A with screw dislocations $(l =1,2,3)$ and $\overline{R}=8$. The region enclosed by a given contour corresponds to $F_{l+1} < F_l$ and a screw dislocation with winding number $l+1$ is energetically favoured as compared to one with  winding number $l$. Thus, in the area $(abcd)$, we have a $l=1$ dislocation while  the area $(dcfe)$ corresponds to $l=2$. The critical twist $\Phi_{ch}^{(c)} = \overline{R}^2/\sqrt 2$ (see text) at which the system becomes cholesteric is $\Phi_{ch}^{(c)} =45.26$.}
\label{fig4}
\end{center}
\end{figure}

For an infinite system there is no mechanism to select 
the number $l$ of screw dislocations. The boundary of a finite Êmesoscopic 
SmA Êintroduces such a mechanism that determines the value of $l$ corresponding to stable screw dislocation solutions as a function of the twist. In a finite system, the order parameter is not equal to $1$ at the boundary, so that 
the boundary conditions used to derive (\ref{bogo}) are not generally satisfied. Hence, the identification of the boundary integral in (\ref{dggbogo}) with the fluxoid (\ref{fluxoid}) is no longer possible and the free energy cannot be minimized just by using the Bogomol'nyi equations (\ref{bogo}). To fix the boundary conditions for a finite system, Êwe impose that the flux of the twist of the SmA through the cylinder is equal to the flux of a cholesteric in the same geometry. To implement it, we consider first the expression of the director $\bs{n}$ in the cholesteric phase \cite{RL88,DGP93} given by 
$ \bs{n}_{ch} (\bs{x})=(0, sin k_0 x, cos k_0 x)$ and obtained by minimizing the cholesteric free energy $F_{ch}=F_{F} - K_2 k_0 \int (\bs{n} \cdot \bs{\nabla} \times \bs{n})$. The flux of the twist through a cylinder of radius $R$ is given by $\pi R^2 \bs{n}_{ch} \cdot \bs{\nabla} \times \bs{n}_{ch} = \pi R^2 k_0$. And because the total flux of the twist of the SmA is $2 \pi R \delta n_{\perp} (R)$, we have
\beq \delta \bs{n}_{\perp} (R)= \delta n_{\perp}(R) \hat{\bs \theta}
=\frac{k_0 R}{2} \hat{\bs \theta} = \frac{
\pi R}{L_p} \hat{\bs \theta}.   \label{bclq} \eeq 
This boundary condition disregards edge dislocations \cite{BKL01,Kamien97}. For the geometry of a cylinder, the current density $\bs{j}$ has only an azimuthal component. We note that on the boundary, $\overline{\bs{\nabla}}_{\perp}\Phi|_{\partial \Omega} = l / \overline{R}$, so that for 
\beq l \le \frac{q_s k_0 R^2}{2} = \Phi_{ch} \label{zerocurrent}\eeq
the current density $j_{\theta}(R) $ is negative. Moreover, since the current density is positive at the 
core of a screw dislocation, there exists a circle along which the current density vanishes.
It allows to split the domain $\Omega$ into two concentric subdomains 
$\Omega = {\Omega_1}
\cup {\Omega_2}$, so that the boundary $\partial \Omega_1$ is the zero current density line. The system can now be 
divided into a bulk ($\Omega_1$) and an edge region ($\Omega_2$). 
We neglect the 
interaction between the screw dislocation 
and the edge current, so that the total free energy density of the system 
is now given by ${\cal F}= {\cal F}(\Omega_1) + {\cal F}(\Omega_2)$. 
Since the boundary $\partial \Omega_1$ is the zero current line, we can approximate ${\cal F}(\Omega_1)$ 
by the Bogomol'nyi 
bound (\ref{inenergy}) \cite{eric99}.  For the edge free energy, we use expression 
(\ref{dlfreeen1}) integrated over $\Omega_2$. The 
twist ${\cal T}(\overline{r}) = \overline{\bs{\nabla}}_{\perp} \times \overline{q}_s \delta \bs{n}_{\perp}$ enters the system up to a distance of order 1 in units of $\lambda_{T} \sqrt2$ \cite{DG72}. Therefore, it decreases from the boundary at $r=R$  with a behaviour  well described by 
\beq {\cal T}(\overline{r}) = \overline{\bs{\nabla}}_{\perp} \times \overline{q}_s \delta \bs{n}_{\perp} = {\Phi_{ch} - l \over  \overline{R} }e^{- (\overline{R}- \overline{r})},  \label{twistom2} \eeq where $\Phi_{ch}$ is defined by (\ref{zerocurrent}). The value at the boundary ${\cal T}(\overline{R})= (\Phi_{ch} - l)/ \overline{R}$ results from the expression of  the total twist $ 2 \pi \Phi_{ch} $ as a sum $2 \pi l + \int_{\Omega_2} {\cal T}(\overline{r})$ of the twists  associated respectively to the dislocations in the subdomain $\Omega_1$ and to the contribution of the subdomain $\Omega_2$. The twist contribution to the free energy in expression (\ref{dlfreeen1}) is thus given by $\int_{\Omega_2} \frac{1}{2}(\overline{\bs{\nabla}}_{\perp} \times \overline{q}_s \delta 
\bs{n}_{\perp})^2 = (l- \Phi_{ch})^2 / 4 \overline{R}$.  Along the same line of arguments, the quantity $\bs{j}/|f|^2 = \overline{\bs{\nabla}}_{\perp} 
\Phi - \overline{q}_s \delta\bs{n}_{\perp}$ is such that 
$\int_{\Omega_2} \left(  (\overline{\bs{\nabla}}_{\perp} \times \bs{j})/|f|^2 + {\cal T}(\overline{r})\right) =0$. Then, $-\bs{j}/|f|^2$ restricted to the subdomain $\Omega_2$  is also given by (\ref{twistom2}).  At the dual point we have $\lambda_T = \xi_{\perp}/ \sqrt2$, so that the amplitude of the 
nematogen density modulation $f$ 
saturates to its equilibrium value 
(undistorted SmA) over a distance $\lambda_T \sqrt{2}$. With the help of these observations, we estimate ${\cal F}(\Omega_2)$ using a variational ansatz, namely that the modulus $|f|$ has a constant value $f_0$ over a ring of width $\lambda_T \sqrt{2}$, included in $\Omega_2$, so that  the term $\overline{\bs{\nabla}}_{\perp}|f||_{\partial \Omega}=0$ \cite{eric99}.  This leads to 
an edge energy of the form:
\bea \int_{\Omega_2}{\cal F} &=&  
{(l- \Phi_{ch})^2 \over 4 \overline{R}} +   \\ &+& \pi \overline{R}  \left( {(l- \Phi_{ch})^2 \over  \overline{R}^2} f_0^2  + (1 - f_0^2)^2 + 2 \alpha^2 f_0^2 \right) . \nonumber \label{edgeen} \eea
We then minimize  with respect to $f_0$, which yields $2(1 - f_0^2) ={(l- \Phi_{ch})^2 \over  \overline{R}^2} + 2 \alpha^2$.  The corresponding  expression for the extremum value of the edge free energy (\ref{edgeen}) together with the bulk term (\ref{inenergy}) gives for the total free energy 
\bea F_{l}(\Phi_{ch}) &= & 2 \pi Êl(1-\alpha^2) + { \pi \over \overline{R}} [{3 \over 2} (l- \Phi_{ch})^2 + 
2 \overline{R}^2\alpha^2] \nonumber \\
&- & {\pi \over 4 \overline{R}^3}
[(l - \Phi_{ch})^2 + 2 \overline{R}^2 \alpha^2 ]^2 . \label{r14} 
\eea 
This relation constitutes the second main result of this letter. In the limit of a radius $R \gg \lambda_T$, the quartic term becomes  negligible compared to the quadratic one, so that the $\Phi_{ch} $-dependence of the total free energy  appears as the envelop of a family of parabolae each indexed by the integer $l$. The system chooses its winding number $l$ in order to minimize the free energy. This provides a relation between the number $l$ of screw dislocations and the twist $\Phi_{ch}$ given by the integer part $l = [\Phi_{ch} - {2 \over 3} \overline{R} (1 - \alpha^2)]$. The entrance of the $(l+1)$-$th$ screw dislocation occurs for a value of the twist, so that $F_{l}=F_{l+1}$. 
The entrance of the first dislocation is thus given by 
$\Phi_{ch} = {1 \over 2} + {2 \over 3} \overline{R}  ( 1 - \alpha^2)$, Ê
which according to (\ref{zerocurrent}) gives for the corresponding twist the expression
$k_0 = {1 \over q_s R}\left( {1 \over R} +  { 4 (1 - \alpha^2)  \over  3 \sqrt{2} \lambda_T }\right)$.
The highest value $k_0^{(c)}$ of the twist that allows the entrance of screw dislocations corresponds to the critical value at which the system becomes cholesteric. It is given by $k_0^{(c)}=\sqrt{\frac{r^2}{gK_2}}$ \cite{RL88} and it corresponds to $\Phi_{ch}^{(c)} = \overline{R}^2/\sqrt 2$. Using this expression in (\ref{zerocurrent}), Êwe obtain a bound for the largest number of screw dislocations that a mesoscopic sample can accomodate. All these results are summarized in Figure \ref{fig3}, which presents the behaviour of the free energy $F_l$ as a function of $\Phi_{ch}$. 
In deriving the expression (\ref{r14}), 
we have implicitely assumed that the parameters $\Phi_{ch} \propto k_0$ and $\alpha^2$ can be changed 
independently. Nevertheless, it has been shown experimentally \cite{GS90}  that both $k_0$ and $\alpha^2$ depend on temperature. 
However, it has been argued that Ê$k_0$ might also be changed at fixed temperature 
\cite{RL88,KS95}. Since we do not know the exact temperature dependence of $k_0$ 
and $\alpha^2$, we have presented in Figure \ref{fig4} the phase diagram 
of a mesoscopic ÊSmA in the $\Phi_{ch}$-$\alpha^2$ plane. It exhibits finite areas where screw dislocations are energetically 
stable.
In conclusion, we have shown that the de Gennes free energy of a smectic-A is characterized by two dimensionless parameters $\kappa$ and the anisotropy $\alpha$. At the dual point $\kappa = 1 / \sqrt 2$, the free energy is topological and it saturates the Bogomol'nyi bound. The role of the boundary of a finite size mesoscopic smectic is to provide a mechanism for the existence of stable screw dislocations. The Bogomol'nyi equations (\ref{bogo}) have been obtained without any specific assumption on the nature of the twist field. Therefore, they share some similarities with those obtained in other situations like mesoscopic superconductors \cite{eric99}, Chern-Simons \cite{jackiw} and Yang-Mills \cite{witten} field theories. These analogies can be pushed further towards other anisotropic layered systems like anisotropic High-$T_c$ superconductors or double layer quantum Hall systems \cite{DLQH}. 

This research was supported in part by the Israel Academy of
Sciences and by the Fund for Promotion of Research at the
Technion.

\end{document}